\documentclass[aps,pra,singlecolumn]{revtex4}
\usepackage{graphicx}
\usepackage{ulem}

\begin{document}


\title{Timing characterization of 100 GHz passively mode-locked discrete mode laser diodes}

\author{David Bitauld, Simon Osborne and Stephen O'Brien}

\affiliation{Tyndall National Institute, University College Cork, Lee Maltings, Cork, Ireland}

\begin{abstract}
We report on the characterization of the timing stability of passively mode-locked discrete mode diode laser
sources. These are edge-emitting devices with a spatially varying refractive index profile for spectral filtering. 
Two devices with a mode-locking frequency of 100 GHz are characterized. The first device is designed to support a 
comb of six modes and generates near Fourier limited 1.9 ps pulses. The second supports four primary modes 
resulting in a sinusoidal modulation of the optical intensity. Using a cross-correlation technique, we measured a 
20 fs pulse to pulse timing jitter for the first device, while, for the second device, a mode-beating (RF) linewidth 
of 1 MHz was measured using heterodyne mixing in a semiconductor optical amplifier. Comparison of these results 
with those obtained for an equivalent Fabry-Perot laser indicates that the spectral filtering mechanism employed 
does not adversely affect the timing properties of these passively mode-locked devices. 
\end{abstract}




\maketitle

\section{Introduction}

Optical data transmission and signal processing are driving increasing interest in the generation of optical 
carriers with specially tailored intensity modulation. Applications such as radio over fiber require a sinusoidal 
modulation of the optical intensity with frequency stability and narrow RF or mode-beating linewidth 
\cite{campany_07}. On the other hand, tailored and highly stable pulse streams are an essential requirement of 
time division multiplexing and optical sampling systems \cite{weber_06, valley}. Optical frequency comb 
generation is a related problem where the creation of a series of equally spaced and mutually coherent spectral 
lines enables coherent communication schemes with ultimate spectral efficiency \cite{delfyett_06, yi_10}.

For microwave frequencies, techniques involving modulation of a continuous wave single mode laser driven by an 
electronic synthesizer can provide optical carriers with narrow linewidth, but external optical 
filtering of a pair of modes is necessary to remove unwanted harmonics \cite{hirata_03, fukushima_03}. 
For higher modulation frequencies, mode-locked laser diodes are promising candidates for pulse stream and frequency 
comb generation \cite{vasilev, avrutin_00, depriest, williams_04, Quinlan_09, hou_2011}. However, in monolithic 
devices, the optical mode spacing, which determines the repetition rate, and the individual modal intensities are 
generally determined by the cavity length and the material gain dispersion, respectively. Hence, tailoring of 
the output spectrum will again require external filtering to adjust modal intensities.
Higher repetition rates have been achieved by harmonic mode-locking techniques including colliding pulse 
mode-locking \cite{chen_91,martins_95, shimizu_97} and in devices based on compound cavities 
\cite{arahira_96,yanson_02}, but these approaches do not in general allow us to address modal intensities 
on an individual basis.

We have developed a simple intracavity spectral filtering technique that selects a finite number of modes in standard 
monolithic Fabry-Perot (FP) diode laser \cite{obrienPN_10}. In these so-called discrete mode (DM) lasers, 
our design approach in principle allows us to fix the harmonic frequency and also to tailor the intensity profile 
of the mode-locked laser output. As examples, we have recently designed and demonstrated two types of DM 
lasers: one producing pulsed output \cite{bitauld_10} and the other a sinusoidal intensity modulation \cite{obrienMTT_10}. 
As timing noise can seriously affect functionalities like data transmission or optical sampling, this paper is dedicated 
to the characterization of the timing noise properties of these devices. The origins of timing noise in mode-locked lasers 
have been extensively studied \cite{Haus_93, Eliyahu, Paschotta}. In the case of semiconductor lasers, spontaneous emission 
noise is the dominant perturbation at high frequency, which is the frequency range we focus on in this work.

Here we present measurements of timing jitter and mode-beating linewidth for two devices with mode-locking frequencies 
of 100 GHz. A natural benchmarking device for this study is a plain Fabry-Perot diode laser with the same cavity length. 
Based on the comparison with the FP device, we conclude  that the spectral filtering that we perform does not adversely 
affect the timing performance of passively mode-locked DM devices. 

This paper is organised as follows: In the next section of this paper we describe the design of each DM
laser that we consider. We also present mode-locked optical spectra and intensity autocorrelation measurements. 
Following this, we describe the timing-jitter measurement, which was performed on the pulsed source and plain 
Fabry-Perot devices using an optical cross-correlation technique. In the penultimate section 
we present measurements of the mode-beating linewidth for the second DM laser. Because of the very large 
frequency involved, this measurement was performed using a heterodyne technique based on nonlinear wave-mixing
in a semiconductor optical amplifier (SOA). Our conclusions and outlook are presented in the final section.
 
\section{Laser design and intensity spectra}

The devices we consider are 875 $\mu$m long monolithic ridge waveguide lasers with a high reflection coating on 
one end and an as-cleaved facet on the other. The active medium is composed of a gain section and a saturable 
absorber section both of which comprise five InP/InGaAlAs quantum wells. A positive current is applied to the gain 
section and a negative voltage is applied to create a saturable absorber section. Three devices will be 
characterized in this work. The first one is a test device with an unperturbed Fabry Perot cavity (FP device) 
while for the two others six and four modes were selected using shallow etched slots to perturb the cavity 
(respectively 6DM and 4DM devices). The precise distribution of slots is determined using an 
inverse scattering approach \cite{obrienPN_10}. 

\begin{figure}
\begin{centering}
\includegraphics[width=12cm]{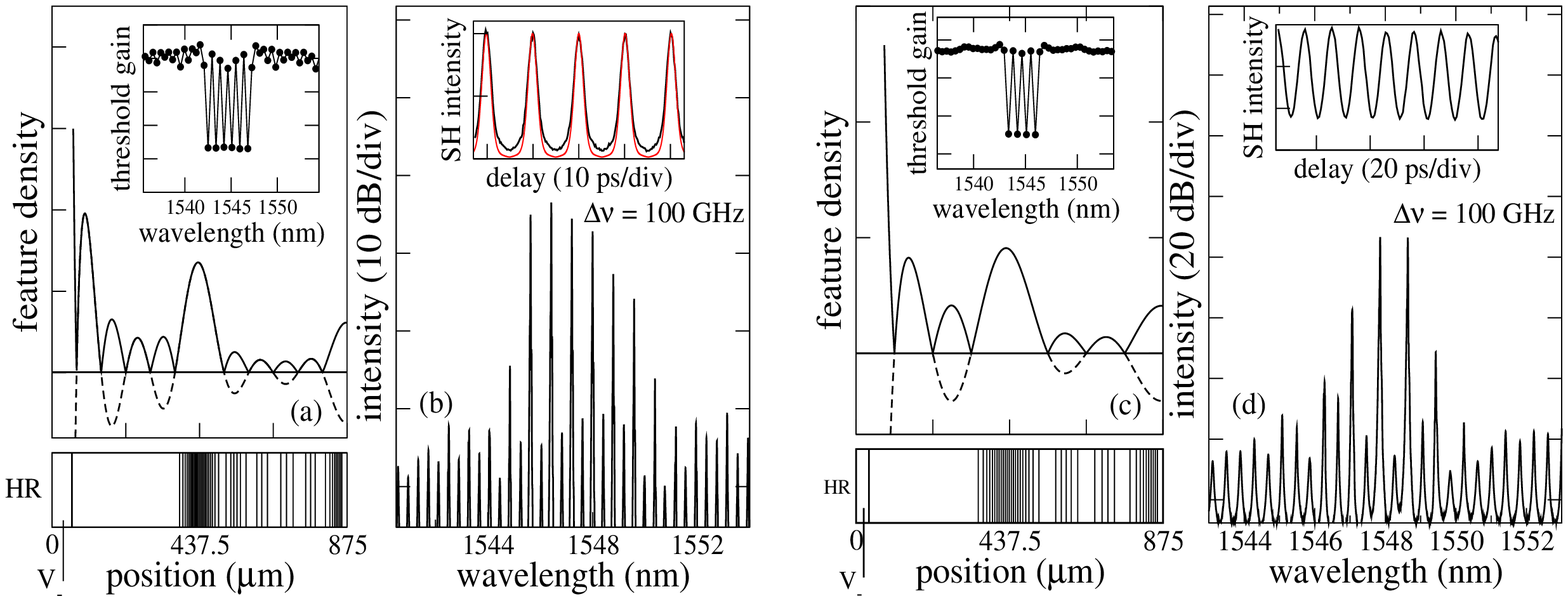}
\caption{\label{fig:figure1} (a) and (c): Feature density functions of the 6DM and the 4DM lasers respectively. 
Dashed lines indicate intervals where the Fourier transform of the spectral filtering function chosen is negative. 
Insets: Calculation of the modal threshold gain for the laser cavities schematically pictured in the lower panels 
of each figure. Lower panels: Laser cavity schematics indicating the locations of the additional features. 
The devices are high-reflection (HR) coated and include a saturable absorber section adjacent
to the HR mirror as indicated. (b): Mode-locked optical spectrum for the 6DM device. The voltage across the 
saturable absorber section is -2.0 V. The current in the gain section is 175 mA. (d)  Mode-locked optical spectrum for
the 4DM device. The voltage across the saturable absorber section is -2.3 V. The current in the gain section is 180 mA
Insets: (b) Autocorrelation measurement (black solid line) and Fourier limited equivalent 
for the 6DM laser (red dashed line) . (d) Autocorrelation measurement for the 4DM laser.}
\end{centering}
\end{figure} 

Schematic pictures of the devices are shown in the lower panels of Fig. 1 (a) and (c). In each case the saturable absorber 
section is placed adjacent to the HR coated mirror. The length of the absorber region was  60 $\mu$m in the case of the 6DM
device, while for the 4DM device, its length was 30 $\mu$m. The density function that determines the cavity geometry 
in each case is shown in Fig. 1 (a) and (c). In both of these examples modes are selected at the second 
harmonic of the cavity which determines a primary mode spacing and mode-locking frequency of 100 GHz. A calculation 
of the modal threshold gain for each device is shown in the insets of Fig. 1 (a) and (c), where we can see that 
six and four non-adjacent primary modes are selected. 

Optical spectra and intensity autocorrelations of the DM devices are shown in  Fig. 1 (b) and (d). 
For each device the value of the gain current and the absorber reverse voltage were chosen to obtain stable 
mode-locking. The parameter values were different in each case because mode locking appears in different 
regions of the phase space. For the 6DM laser the gain current was 175 mA and the absorber voltage was -2.0 V.  
From the optical spectrum of the 6DM device we can see that six modes are significantly enhanced with respect 
to the background modes. There is also some asymmetry of the optical spectrum which we attribute to a mismatch 
between the position of the gain peak and the frequency of the selected modes. From the intensity autocorrelation 
of the 6DM laser shown in the inset of Fig. 1 (b), we can see that well developed pulses are generated at 100 GHz 
repetition rate. The red line in this case corresponds to the Fourier limited pulse train autocorrelation 
we would expect from the measured optical spectrum where only the ten central peaks were taken 
into account.  Assuming Gaussian shaped pulses, the deconvolved full width at half maximum (FWHM) pulse duration 
is obtained by dividing the autocorrelation FWHM by $\sqrt{2}$. We find that in this case the pulse duration is 
1.90 ps with a very similar Fourier limited value. For comparison, the FP laser (not displayed here) generated pulsed
output with a minimum pulse duration of 1.56 ps when the gain current was 130 mA and the absorber voltage 
was -1.6 V. In this case the corresponding Fourier limited value was 0.86 ps, which indicates the presence of 
significant pulse chirp or variation of the central wavelength leading to a broadening of the optical spectrum
measured over many pulses. We conclude that although the 6DM laser generated pulses of longer duration on account of 
the limited number of pre-selected modes, unlike the FP laser, the generated pulses are close to the Fourier limit 
in the DM device. 

In the case of the 4DM device, the optical spectrum shown was obtained with a gain section current of 180 mA and an 
absorber voltage of -2.3 V. The autocorrelation shown in the inset in this case is characteristic of a beating between 
two modes of approximately equal amplitude. Note that for a purely sinusoidal mode-beating, we would expect the ratio 
of the maxima and minima of the autocorrelation to be equal to 3. Here the  ratio is somewhat larger than 
this value, which is due to the presence of the pair of satellite modes that add a small harmonic component to the pure 
sinusoidal waveform. Although the presence of these satellite modes represents an inconvenience in this case, they are 
necessary to overcome wavelength bistability by enabling mode-locking of the device \cite{obrienMTT_10}. It's 
interesting to note that our design approach will in principle allow us to adjust the losses of these modes independently 
from the central pair, which may allow us to obtain an intensity modulation which is closer to the ideal limit. This
freedom to individually address the linear losses of each of the selected modes may also enable us to create coherent 
combs of modes with approximately uniform intensity for applications in coherent communications.

The output powers of the considered devices varied from 18.0 mW in the case of the Fabry-Perot laser, to 7.4 mW in the case 
of the 4DM device, and 2.2 mW in the case of the 6DM. We attribute the discrepency between the output powers of the
DM devices to be due to the fact that the 6DM device included a longer saturable absorber section, while the reduced
power of both of these devices relative to the FP laser is likely due to the etched features, which inevitably introduce 
additional scattering losses. 

\section{Timing jitter}

Techniques that extract the laser intensity phase noise from a measurement of the intensity power spectrum 
\cite{vd_Linde_86} have become the most common way to characterize timing stability in mode-locked lasers. 
However, these techniques require the use  
of photodiodes and spectrum analyzers with a bandwidth significantly higher than the repetition rate of the signal. 
The devices that we consider here have a repetition rate of 100 GHz which makes this technique inapplicable.
On another hand, optical intensity cross correlation (XC) techniques have been proposed \cite{Haus_02, Tourrenc} 
that allow us to measure fast timing variations. This type of measurement is sensitive to fluctuations faster than 
$(4\tau)^{-1}$, where $\tau$ is the delay used in the XC (up to approximately 100 ns in this paper). This means that 
this measurement is typically sensitive to intrinsic noise sources related to spontaneous emission but not to 
environmental and other extrinsic noise sources that are usually relevant at lower frequencies.

\begin{figure}[t]
\begin{centering}
\includegraphics[width=12cm]{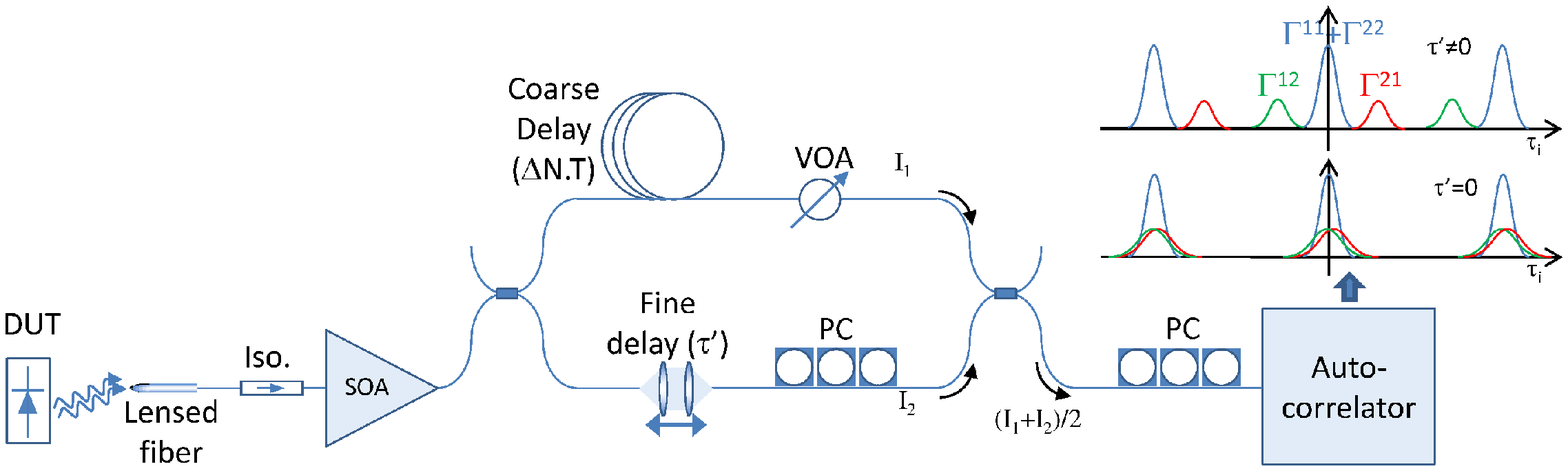}
\caption{ \label{fig:JitterSetup} Optical cross-correlation setup. The top right inset is a schematic representation 
of the autocorrelation result in a general case ($\tau'\neq 0$) and when the delay between the two arms is an integer 
number of periods ($\tau =0$).  $\Gamma^{11}$ and $\Gamma^{22}$ are the autocorrelations of the signals that propagated 
in each arm and  $\Gamma^{12}= \Gamma^{12}_{\Delta N}(\tau^{'} + \tau_{i})$ and  $\Gamma^{21}= \Gamma^{12}_{\Delta N}(\tau^{'} - 
\tau_{i})$ are the cross-correlation of the two signals with a coarse delay of $\Delta N$ periods. }
\end{centering}
\end{figure} 

The XC we are dealing with here is in fact an autocorrelation of the intensity pulse train involving two different pulses 
from the same train separated by the time delay of the measurement. The statistical distribution associated with the time 
between two pulses results in a broadening of the XC compared to the autocorrelation of a single pulse. Indeed, the XC is a 
convolution of the autocorrelation with the probability density function of the time between two pulses (appendix eq. \ref{GDN}). 
Assuming Gaussian processes, the timing jitter's standard deviation can be extracted from this measurement.
Experimentally, optical autocorrelations and XCs are performed by splitting 
the light beam and having the two resulting beams propagate different distances resulting in a delay $\tau$. Then they 
are recombined in a nonlinear ($\chi^{2}$) crystal. If the two beams are collinear the generated second harmonic intensity 
$I_{\mbox{\scriptsize SH}}$ is related to the autocorrelation by the relation
\begin{eqnarray}
\langle I_{\mbox{\scriptsize SH}}(t)\rangle_t \propto \langle \left(I\left(t\right) + I\left(t + \tau\right)\right)^2 \rangle_t  
\propto \langle I(t)^2\rangle_t + \langle I(t) \cdot I(t + \tau) \rangle_t .
\end{eqnarray}
In the above $I(t)$ is the pulse train intensity and $\langle \cdots \rangle_{t}$ indicates average over time.  We can see 
that the autocorrelation (or XC) term $\langle I(t) \cdot I(t + \tau) \rangle_t$ is added to a background $\langle I(t)\rangle_t$ 
independent of $\tau$. However, a background free autocorrelation can be obtained in non-collinear configurations.
Commercial autocorrelators usually are background free but their available range of delay is not sufficient to measure jitter. 
A background free cross-correlation setup can then be built with free space elements as in \cite{Haus_02}. It can also be 
implemented with fiber optic elements as in \cite{Tourrenc} by splitting and recombining the beams with fiber beam 
splitters before illuminating the non-linear crystal. The latter technique is much easier to set-up but it has the disadvantage
that it is not background free. We will show later that this drawback can seriously impede the quality of the result when the pulse width 
and the jitter distribution width are not negligible compared to the repetition rate. What is more, the manual optical 
delay makes the measurement very slow. Therefore it is very sensitive to environmental conditions causing a slow drift in the signal 
period and the optical delay, which can artificially broaden the XC. Here we propose a slightly different technique where we perform 
the (fast) autocorrelation of the recombined beam with a commercial background free autocorrelator.

Fig. \ref{fig:JitterSetup} displays the experimental setup with a schematic representation of the measurement obtained.
We can show (appendix eq. \ref{Gi}) that the  XC appears in the data obtained but it is added to autocorrelation terms. 
The measurement is given by:
\begin{eqnarray}
 \Gamma_i(\tau_{i}) = \Gamma^{11}(\tau_{i}) + \Gamma^{22}(\tau_{i}) 
+ \Gamma^{12}_{\Delta N}(\tau^{'} + \tau_{i}) + \Gamma^{12}_{\Delta N}(\tau^{'} - \tau_{i}),
\end{eqnarray}
$\Gamma^{11}$ and $\Gamma^{22}$ being the autocorrelations of the signals that propagated in each arm and  
$\Gamma^{12}_{\Delta N}$ the cross-correlation of two pulses separated by $\Delta N$ periods. $\tau_{i}$ is the internal delay 
of the autocorrelator and $\tau^{'}$ is the external fine optical delay. $\Gamma^{11}$ and $\Gamma^{22}$ are centered on 
$\tau_{i} = 0$ by definition of an autocorrelation. The two terms can be slightly different if dispersion broadening is 
significant.

\begin{figure}[t]
\begin{centering}
\centering\includegraphics[width=10cm]{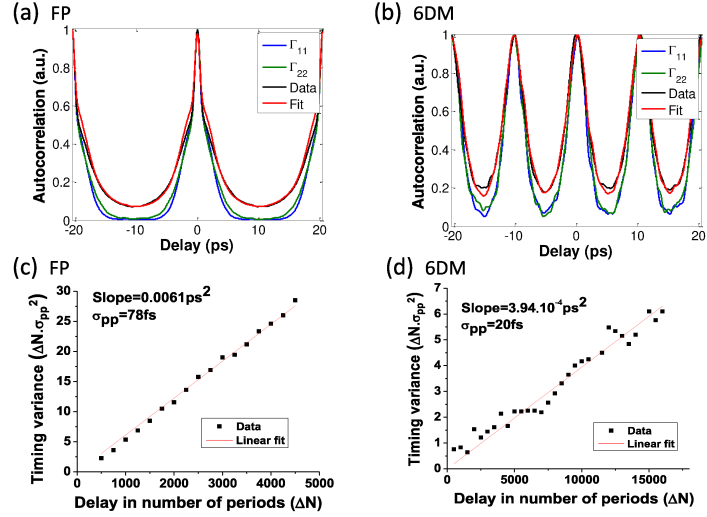}
\caption { \label{fig:JitterPlots} Measurement of the timing jitter. (a) and (b) display the autocorrelation measurements and 
data fittings performed with the FP device at 16~m delay ($\Delta N=4000$) and the 6DM device at 24~m delay  ($\Delta N=12000$)  
respectively. $\Gamma^{11}$ and $\Gamma^{22}$ are the autocorrelation of the intensity from each arm separately. "Data" is the
measurement of $\Gamma_i$ and "Fit" is the fitted theoretical curve of $\Gamma_i$. In (c) and (d) the black squares represent 
the timing variances used to fit the $\Gamma_i$ curves for different delays and the red line is a linear fit of those points.}
\end{centering}
\end{figure} 

The two XC terms are off-centered except in the case where the delay between the two arms is a multiple of the signal 
period. If the pulses are well-separated as in the top right inset of Fig. \ref{fig:JitterSetup}, the XCs can be 
isolated from the autocorrelations and measured directly. Since the devices we characterize here have a small number 
of cavity modes, the pulses are not well-separated enough to
isolate the XCs directly. In order to retrieve the timing-jitter from the measurement we have to fit the expected shape 
of the autocorrelation to the measurement with the jitter as a fitting parameter.  $\Gamma^{11}$ and $\Gamma^{22}$ can be
measured separately by switching off one arm or the other. The noiseless  XCs is obtained by assuming that 
the effect of dispersion, if significant, can be approximated as a convolution with a Gaussian $G_{D}$. The width of 
$G_{D}$ is determined by fitting $\Gamma^{22}*G_{D}$ to $\Gamma^{11}$. Therefore, the expected shape of the XC is  
$\Gamma^{22}*G_{D}*G_{\Delta N}$, where the width of the jitter Gaussian $G_{\Delta N}$ is the only unknown. The external fine 
delay $\tau^{'}$ is tuned so that the total delay between the two arms is a multiple of the period and both XCs overlap 
around $\tau_{i} = 0$ as in the lower right inset of Fig. \ref{fig:JitterSetup}. If we assume a symmetrical shape for the 
pulses the XCs become identical. A variable optical attenuator (VOA) is used to balance the intensity in both arms. 
Thus the function we fit to the measurement is:
\begin{eqnarray}
\Gamma_i(\tau_{i}) = \Gamma^{11}(\tau_{i}) + \Gamma^{22}(\tau_{i}) + 2\times \Gamma^{22}(\tau_{i})*G_{D}(\tau_{i})*G_{\Delta N}(\tau_{i}).
\end{eqnarray}
The width of $G_{\Delta N}$, i.e. the jitter standard deviation $\sqrt{\Delta N} \sigma_{pp}$, is the only fitting parameter.
Fig. \ref{fig:JitterPlots} (a) and (b) display, for a Fabry-Perot and for a 6 mode device, the measured and fitted curves 
as well as the autocorrelation of the intensity in both arms. We can see that, as our pulses are very close to each other, 
the diminution of the extinction ratio caused by the jitter is more visible than the broadening of the peaks. 
This is why it is very important to have a well-defined zero and a background free measurement makes this very 
convenient.

Fig. \ref{fig:JitterPlots} (c) and (d) display the timing variances measured for different coarse delays. As expected for a 
white noise pulse-to-pulse jitter, the variance increases linearly as a function of delay with a slope equal to the 
square of the pulse-to-pulse jitter standard deviation $\sigma_{pp}$. A linear fit of the data allows us to retrieve 
the value of $\sigma_{pp}$. It is 78 fs for the FP device and 20 fs for the 6 mode device. Because of the difference in 
repetition rate, the timing noise of the FP has to be compared with that of the DM after two periods, which corresponds 
to a single round trip of the intracavity pulse in both FP and DM. This gives us a jitter of $20\times \sqrt{2}=28$~fs,
which is significantly less than the value of 78~fs that was measured for the FP. 
This shows that the patterning of the laser cavity for mode selection does not adversely affect the timing stability of the 
device. In fact, accounting for the difference in repetition rate, the timing stability is seen to have improved in the 6DM
device.

\section{RF linewidth}

The device with four pre-selected modes that we consider was designed to generate an optical carrier for millimeter-wave 
signals. This device represents a minimal mode-locked system for such an application, and can be regarded as integrating the
functions of passive mode-locking and spectral filtering in a single monolithic device \cite{obrienMTT_10}. Because in this 
case the intensity profile is sinusoidal, the relevant characterization of the timing noise is given by a measurement of 
the RF or mode-beating linewidth of the signal. As was the case for the timing jitter of pulsed sources, a direct measurement 
of the intensity is not possible at high repetition rates. A technique to overcome this problem has been recently proposed 
\cite{Latkowski}. The approach is to produce a beating between two single mode tunable lasers and to mix this signal in a 
semiconductor optical amplifier (SOA) together with the optical beat signal to be measured. Non-linearities 
due to spectral hole burning and carrier heating generate a slight modulation of the output at a frequency corresponding to 
the difference between the measured signal and the beating of the single-mode lasers. This modulation is slow enough to be 
measured with conventional photodiodes. Here we use a similar setup except we do not amplify our signal before mixing it in the 
SOA.

\begin{figure}[htbp]
\begin{centering}
\centering\includegraphics[width=9cm]{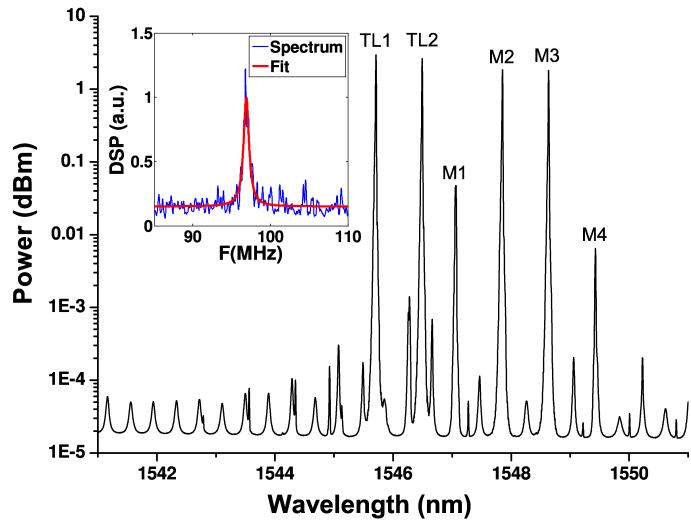}
\caption { \label{fig:OSAPSD} Optical spectrum at the output of the SOA. Inset: Electrical spectrum at the output of the photodiode 
(narrow blue line), and Lorentzian fit (thick red line). }
\end{centering}
\end{figure} 

An optical spectrum of the output of the SOA is displayed in Fig. \ref{fig:OSAPSD}. In the right half of the figure we can see 
the four modes of the device (two main modes M2 and M3 and two satellites M1 and M4) and on the left we can see the two tunable 
lasers TL1 and TL2. The wavelength of the tunable lasers are chosen to be half-way between residual modes of the DM device 
in order to avoid interference. The intensity is measured with a fast photodiode connected to an oscilloscope. A series of 50 
$\mu$s time traces is recorded by the oscilloscope before being analyzed. The length of the time traces has been chosen so that 
the wavelengths of the tunable lasers do not have time to drift during the measurement. The Fourier transform of these time traces 
is fitted with a Lorentzian function as shown in the top left inset of Fig. \ref{fig:OSAPSD}. The spectrum data displayed 
have been smoothed for clarity. The accuracy of this measurement is limited by the stability of the two tunable lasers. This was
determined to be $\sim$ 200~kHz from a measurement of beating linewidth of their outputs over a 50~$\mu$s interval.

The RF linewidth of the 4DM device has been measured for different parameters ranging from 115 to 180~mA for the gain current and 
-1.8 to -2.5~V for the absorber section. Its value varied from 1 to 1.8~MHz. For comparison we also performed this measurement on the 
6DM and the FP lasers and obtained values for the RF linewidth of 3~MHz and 4.5~MHz respectively. Assuming white Gaussian noise 
these two values can be related to the jitter values measured in the previous section through the relation 
$\Delta\nu_M=\frac {2\pi(M \sigma_{pp})2}{T^3}$ (appendix eq. \ref{S}). The jitter measurements of the 6DM and FP lasers predict 
RF linewidths of 2.5~MHz and 4.8 MHz respectively, which are in reasonable agreement with the directly measured values. 
The narrow RF  linewidth of the DM lasers confirms their good frequency stability compared to the equivalent FP. We note also 
the fact that because the intensity modulation is produced by two modes which share the same cavity, we can obtain 
a mode-beating linewidth that is much smaller than the optical linewidths of the modes. The optical linewidths of the 
selected modes in the 6DM device were measured to be around 40 MHz in the mode-locked state. Although this value is too 
large to make the current DM devices of interest as frequency comb sources, we note that the corresponding optical linewidths
in the FP device were more than an order of magnitude larger again. Thus intracavity spectral filtering such as we 
perform here may be advantageous from the point of view of frequency comb generation using mode-locked diode lasers.  

\section{Conclusion}

In conclusion, discrete mode lasers with engineered spectra are a promising way to produce optical 
carriers with tailored intensity profiles. We showed that they are closer to the Fourier limit and have 
comparable or better timing fluctuations than their FP equivalents fabricated with the same material. A possible
explanation for the latter improvement may be that spectral filtering limits dispersion coupled timing fluctuations 
(so-called Gordon-Haus jitter \cite{gordon}).  We expect their performance can be further improved by optimizing 
the location of the spectral filter with respect to the material gain dispersion, and by optimising the gain and 
absorber material and the heterostructure design for stable mode-locking. 

\section*{Acknowledgments}

This work was supported by Science Foundation Ireland and Enterprise Ireland. 
The authors thank Eblana Photonics for the preparation of sample devices. 

\begin{center}
{\bf APPENDIX}
\end{center}

\appendix

In order to retrieve the timing jitter from the optical XC measurement, pulses are usually considered Gaussian so that the 
squares of the pulse width and timing standard deviation are additive. The technique we use here is based on the fitting of the data 
and the Gaussian approximation was not always sufficient. For this reason we include the pulse shape or more 
exactly its autocorrelation in the calculation. Besides, given the deliberately limited number of optical modes supported by our lasers, 
the pulses tend to overlap in time. A Fourier series representation of the intensity signal is then more 
appropriate. In this appendix we derive the theoretical effects of the noise on the XC as it applies to our particular 
measurement.

During its propagation in the laser cavity, the pulse is constantly subjected to amplitude and timing fluctuations due 
to spontaneous emission. In the case of passive mode-locking, the deviation from the average position in time $J(t)$ 
drifts without bound integrating over time the white noise timing fluctuations.  $J(t)$ is therefore a random walk. We can 
define \mbox{$J_{\tau}(t)\equiv J(t+\tau)-J(t)$} as the timing change accumulated over a time $\tau$ with a 
linearly increasing variance  $<J_{\tau} (t)^2 >_t= \frac{\tau}{T}\sigma_{pp}^2$. Here $\sigma_{pp}$ is the round 
trip time standard deviation, and $T$ is the average round trip time.

The  noiseless intensity pulse train $I_0(t)$ decomposes in Fourier series as
\begin{eqnarray}
I_0(t)=  \frac{1}{T}   \sum_M     \tilde {p}   \small{   \left(   \scriptstyle{ \frac{M}{T}}  \right) }   e^{2i\pi\frac{M}{T}t}
\end{eqnarray}
where $\tilde {p}(\nu)= \mathcal{ {\Large F}} \Big[ p(t) \Big] (\nu)$ is the Fourier transform of the pulse shape. The noisy 
intensity is $I(t)=I_0\big( t+J(t)  \big)$. The autocorrelation of $I(t)$ yields
\begin{eqnarray}
\Gamma (\tau) = \left< I(t)I(t+\tau) \right>_t=     \frac{1}{T^2}     \sum_M       \left|  \tilde {p}   \left(\scriptstyle{ \frac{M}{T}}\right)  \right|^2 
                e^{  2i\pi \frac{M}{T} \tau  }     \left<         e^{  2i\pi \frac{M}{T}J_{\tau}(t) }                         \right>_t
\end{eqnarray}
$J_{\tau}(t)$ being a Gaussian process we have  $\left<         e^{  2i\pi \scriptstyle{ \frac{M}{T}}J_{\tau}(t) }                         \right>_t =
								e^{     -\frac{1}{2}    \left< \left|2\pi \frac{M}{T} J_{\tau}(t) \right|^2\right>_t }  $, therefore
\begin{eqnarray} \label{G}
\Gamma (\tau)    &=&     \frac{1}{T^2}     \sum_M       \left|  \tilde {p}   \left(\scriptstyle{ \frac{M}{T}}\right)  \right|^2 
                e^{  2i\pi \frac{M}{T} \tau  }          e^{  - \frac{2}{T^3}(\pi M\sigma_{pp})^2|\tau|  }                     
\end{eqnarray}
In order to simplify this formulation we write the delay as the sum of a coarse component $ \Delta N \cdot T$, where $\Delta N$ 
is the integer number of periods contained in the delay, and a fine component $\tau'$ so that $\tau=\Delta N \cdot T+\tau'$.
The XC can then be expressed as a function of $\tau'$ around a coarse delay $\Delta N \cdot T$:
\begin{eqnarray}
\Gamma_{\Delta N} (\tau ')    &=&     \frac{1}{T^2}     \sum_M       \left|  \tilde {p}   \left(\scriptstyle{ \frac{M}{T}}\right)  \right|^2 
                e^{  2i\pi M\cdot\Delta N  }        e^{  2i\pi \frac{M}{T} \tau '  }          e^{  - \frac{2}{T^3}(\pi M\sigma_{pp})^2|\Delta NT+\tau '|  }                     
\end{eqnarray}
The first exponential is always equal to $1$ because $M\cdot\Delta N$ is integer while, in the third exponential, the $\tau'$ dependence 
can be neglected. Therefore we can 
write the XC around a coarse delay $\Delta N\cdot T$ as a convolution of the XC without noise $\Gamma_0$ and the 
Gaussian distribution of the timing deviation between two pulses separated by $\Delta N$ periods:
\begin{eqnarray}\label{GDN}
\Gamma_{\Delta N} (\tau ')    &=&   \Gamma_0 (\tau ')    * G_{\Delta N}(\tau ')                     
\end{eqnarray}
with
\begin{eqnarray}
 \Gamma_0 (\tau ')  = \left< I_0(t)I_0(t+\tau') \right>_t=  \frac{1}{T^2}     \sum_M       \left|  \tilde {p}   
\left( \scriptstyle{ \frac{M}{T}}  \right)  \right|^2   e^{ 2i\pi \frac{M}{T} \tau '  }                 \nonumber    
\end{eqnarray}
and
\begin{eqnarray}
G_{\Delta N}(\tau ')   =      \mathcal{ {\Large F}} ^{-1}   \Big[ e^{  -2 (\pi \nu \sigma_{pp})^2\Delta N } \Big](\tau')=\frac {1}{\sqrt{2\pi\Delta N}\sigma_{pp}} 
                                           \exp \Bigg(      - \frac{1}{2} \left(\frac{\tau '}{\sqrt{\Delta N}\sigma_{pp} } \right)^2  \Bigg).  \nonumber
\end{eqnarray}
Experimentally, if the delay $\tau$ results from propagation in different lengths of dispersive media the pulses in each arm will experience
different broadenings. Thus we define $I_1$ and $I_2$ to be the intensities of the pulse streams that propagated in each arm of the cross-correlator and  
$p_1$ and $p_2$ to be the intensity profiles of the individual pulses composing those streams. The XC without jitter then becomes:
\begin{eqnarray}
\Gamma_{0}^{12}(\tau') =
	  \frac{1}{T^2}     \sum_M       \tilde {p}_1\left(\scriptstyle{ \frac{M}{T}}\right)^{*}  \tilde {p}_2\left(\scriptstyle{ \frac{M}{T}}\right)     e^{  2i\pi \frac{M}{T} \tau '  }    
\end{eqnarray}
where  $^*$ stands for complex conjugate. Following the same steps that led to equation \ref{GDN} we derive the cross-correlation including jitter:
\begin{eqnarray}\label{GDN12}
\Gamma_{\Delta N}^{12}(\tau')=\Gamma_{0}^{12}(\tau') * G_{\Delta N}
\end{eqnarray}
This is the standard cross-correlation measured in \cite{Haus_02} and \cite{Tourrenc} for a coarse delay $\Delta N\cdot T$ and a fine delay $\tau'$.

In our setup we measure the autocorrelation of the signal $I_1+I_2$ as a function of the internal delay of the autocorrelator $\tau_i$, while 
the usual technique measures the cross-correlation of $I_1$ and $I_2$ as a function of the fine delay $\tau'$. Here the autocorrelation measured 
by the autocorrelator is:
\begin{eqnarray}\label{Gi}
& \Gamma_i(\tau_i) = \Bigg<      \Big( I_{1}(t)+I_{2}(t+\tau')  \Big)  .     \Big( I_{1}(t+\tau_i)+I_{2}(t+\tau'+\tau_i)  \Big)                   \Bigg>_t     \nonumber\\  
&	= \Bigg<      I_{1}(t)I_{1}(t+\tau_i)  +I_{2}(t+\tau')I_{2}(t+\tau'+\tau_i)   +     I_{1}(t)I_{2}(t+\tau'+\tau_i)  +I_{2}(t+\tau')I_{1}(t+\tau_i) \Bigg>_t \nonumber\\
& = \Gamma^{11}(\tau_i) + \Gamma^{22}(\tau_i) + \Gamma_{\Delta N}^{12}(\tau'+\tau_i) +\Gamma_{\Delta N}^{12}(\tau'-\tau_i),
\end{eqnarray}
where $\Gamma^{11}$ and $\Gamma^{22}$ are the autocorrelations of $I_1$ and $I_2$ respectively and $ \Gamma_{\Delta N}^{12}$ is defined by equation \ref{GDN12}.

The power spectrum can be found by performing the Fourier transform of equation \ref{G}:
\begin{eqnarray}\label{S}
S(\nu)= \mathcal{ {\Large F}} \left[\Gamma (\tau)\right]  (\nu)  &=&     \frac{1}{T^2}     \sum_M     \left|  \tilde {p}   
\left(\scriptstyle{ \frac{M}{T}}\right)  \right|^2 
               \mathcal{ {\Large F}} \left[    e^{  2i\pi \frac{M}{T} \tau  }          e^{  - \frac{2}{T^3}(\pi M\sigma_{pp})^2|\tau|  }                  \right]   (\nu)   \nonumber
\\
&=& \frac{1}{T^2}     \sum_M       \left|  \tilde {p}   \left(\scriptstyle{ \frac{M}{T}}\right)  \right|^2   \delta \left(\nu-\frac{M}{T}\right)*
	 \frac{1}{  \frac{ (\pi M \sigma_{pp})^2}{T^3} + \frac{ T^3 \nu^2}{(\pi M \sigma_{pp})^2}  }.
\end{eqnarray}
From this expression, we see that each mode M is a Lorentzian of FWHM linewidth: $\Delta\nu_M=\frac {2\pi(M \sigma_{pp})^2}{T^3}$.\\

\end{document}